\documentclass[prd,aps,nofootinbib,showpacs,preprintnumbers,amsmath,amssymb,floatfix]{revtex4}

\usepackage{times}
\usepackage{hyperref}

\usepackage{graphicx}
\usepackage{dcolumn}
\usepackage{bm}
\usepackage{slashed}

\begin{document}

\title{Sum rules for leading vector form factors in hyperon semileptonic decays}

\author{
Rub\'en Flores-Mendieta
}
\affiliation{
Instituto de F{\'\i}sica, Universidad Aut\'onoma de San Luis Potos{\'\i}, \'Alvaro Obreg\'on 64, Zona Centro, San Luis Potos{\'\i}, S.L.P.\ 78000, Mexico
}

\author{
Roberto Padr\'on-Stevens
}
\affiliation{
Instituto de F{\'\i}sica, Universidad Aut\'onoma de San Luis Potos{\'\i}, \'Alvaro Obreg\'on 64, Zona Centro, San Luis Potos{\'\i}, S.L.P.\ 78000, Mexico
}

\date{\today}

\begin{abstract}
By considering that the weak currents and the electromagnetic current are members of the same $SU(3)$ octet, two sum rules involving leading vector form factors in hyperon semileptonic decays are derived in the limit of exact flavor $SU(3)$ symmetry. Deviations from this limit arise from second-order $SU(3)$ symmetry-breaking effects, according to the Ademollo-Gatto theorem. The $1/N_c$ expansion of QCD is used to evaluate such effects. One sum rule vanishes identically even in the presence of symmetry breaking and the other one obtains contributions mainly from the $\mathbf{10}+\overline{\mathbf{10}}$ representation. Results obtained in (heavy) baryon chiral perturbation theory are used to test the validity of these sum rules. To order $\mathcal{O}(p^2)$ in the chiral expansion, results are encouraging.
\end{abstract}


\maketitle

\section{Introduction}

In 1963, N.\ Cabibbo proposed a model for weak hadronic currents based on $SU(3)$ symmetry \cite{cab63}. The model relied on the validity of the $V-A$ theory, the conserved vector current (CVC) hypothesis and, in order to preserve the universality of the weak interactions, introduced $\theta_c$---the Cabibbo angle---which must be determined experimentally. With the advent of the standard model of quarks and leptons and their interactions, in present-day terminology, the hadronic weak current can be expressed directly in terms of quark fields and $\theta_c$ is the rotation angle between the first two generations and the only parameter relevant to hadronic physics involving the light quark sector.

Even from its conception, the Cabibbo model has been a key approach to describing hyperon semileptonic decays (HSD). It should be kept in mind that the model has never been intended to be exact since $SU(3)$ is a broken symmetry. For half a century, the departure of the exact symmetry limit has been scrutinized using several methods in order to find discrepancies between theory and experiment; the leading weak form factors in HSD are the usual probes (see Ref.~\cite{fmg} and references therein for a brief description about some methods used so far).

In particular, according to the Ademollo-Gatto theorem \cite{ag}, the leading vector form factors $f_1$ are protected against $SU(3)$ symmetry-breaking (SB) corrections to lowest order in $m_s-\hat{m}$, where $\hat{m}$ denotes the mean mass of the up and down quarks.

The purpose of the present paper is not to provide evidence of SB in $f_1$ {\it per se} but rather, using general properties shared by the electromagnetic and weak currents as members of an $SU(3)$ octet, to propose two sum rules involving $f_1$ which are valid in the symmetry limit. Departures from this limit are then evaluated using the $1/N_c$ expansion of QCD so the modified sum rules are provided to second order in SB. The analysis allows us to identify the flavor representations responsible for SB in the sum rules.

This paper is organized as follows. In Sec.~\ref{sec:srsu3} the two sum rules involving leading vector form factors are derived by taking the matrix element of the vector current between the baryons within a $V=1$ multiplet. The resulting expressions are valid in the exact $SU(3)$ symmetry limit. In the presence of SB, two modified expressions are then provided. In Sec.~\ref{sec:add} a brief discussion about the procedure used by M.\ Ademollo and R.\ Gatto is provided, followed by a survey on the $1/N_c$ expansion of QCD in Sec.~\ref{sec:sur}. In Sec.~\ref{sec:bff} the $1/N_c$ expansion for the baryon vector current including first- and second-order SB is constructed. Restrictions imposed in the nonrenormalization of the baryon electric charge fix several operator coefficients which also participate in $f_1$. Consequently, a few of them survive and are the ones which produce SB to second order. The sum rules are tested with some analytical results obtained in the framework of (heavy) baryon chiral perturbation theory and the results are encouraging. Some closing remarks are listed in Sec.~\ref{sec:sum}.

\section{\label{sec:srsu3}Sum rules for baryon vector form factors}

Motivated by the success of the Gell-Mann--Okubo formula for baryon octet masses, T.N.\ Pham observed that the matrix elements of the $V=1$ $V$-spin multiplet can be related to each other in the exact flavor $SU(3)$ symmetry limit \cite{pham}. 
Starting from the two $I$-spin relations,
\begin{subequations}
\begin{equation}
\langle I=1,I_3=0 | \overline{u} \Gamma^n d | I=1,I_3=1 \rangle = -\langle I=1,I_3=-1 | \overline{u} \Gamma^n d | I=1,I_3=0 \rangle,
\end{equation}
and
\begin{equation}
\langle I=0,I_3=0 | \overline{u} \Gamma^n d | I=1,I_3=1 \rangle = \langle I=1,I_3=-1 | \overline{u} \Gamma^n d | I=0,I_3=0 \rangle,
\end{equation}
\end{subequations}
the rotated $V$-spin versions of the above relations read
\begin{subequations}
\label{eq:vspin}
\begin{equation}
\langle V=1,V_3=0| \overline{u} \Gamma^n s|V=1,V_3=1\rangle = -\langle V=1,V_3=-1| \overline{u} \Gamma^n s|V=1,V_3=0\rangle,
\end{equation}
and
\begin{equation}
\langle V=0,V_3=0| \overline{u} \Gamma^n s|V=1,V_3=1\rangle = \langle V=1,V_3=-1| \overline{u} \Gamma^n s|V=0,V_3=0\rangle,
\end{equation}
\end{subequations}
respectively, where the bilinear forms $\overline{q}_1 \Gamma^n q_2$ are given in terms of quark fields $q_i$ and the matrices $\Gamma_{\alpha\beta}^n$ are the ones which appear in applications of the Dirac theory, namely, $\Gamma^S=\openone$, $\Gamma_\mu^V=\gamma_\mu$, $\Gamma_\mu^A = \gamma_5\gamma_\mu$ and so on \cite{bd}.

The decomposition of the $SU(3)$ octet into eigenfunctions of $V^2$ read
\begin{eqnarray}
&  & |\Xi^-\rangle = |V=1,V_3=1 \rangle, \\
&  & \left| \frac12 \Sigma^0 + \frac{\sqrt{3}}{2} \Lambda \right\rangle = |V=1,V_3=0 \rangle, \\
&  & |p\rangle = |V=1,V_3=-1 \rangle, \\
&  & \left| \frac{\sqrt{3}}{2} \Sigma^0 - \frac12 \Lambda \right \rangle = |V=0,V_3=0 \rangle.
\end{eqnarray}

Relations (\ref{eq:vspin}) apply to the matrix elements of any $SU(3)$ octet $\Delta S=1$ operator. For instance, the GMO relation is straightforwardly obtained from Eq.~(\ref{eq:vspin}a) by relating $\langle B_2|\partial_\mu(\overline{u} \gamma_\mu s)|B_1\rangle$ to $[f_1^{SU(3)}]_{B_1B_2}(M_{B_2}-M_{B_1})$, where $f_1^{SU(3)}$ stands for the $SU(3)$ symmetric value of the leading vector form factor $f_1$ at zero recoil and $M_{B_i}$ is the mass of baryon $B_i$ \cite{pham}. Similarly, for $\Gamma^n=\Gamma_\mu^A$ some interesting relations for the axial-vector to vector form factor ratios $g_1/f_1$ can also be found \cite{pham}.

The analysis can be extended to the vector current by using $\Gamma^n=\Gamma_\mu^V$. As a result, two simple although nontrivial expressions are obtained, namely,
\begin{subequations}
\begin{equation}
\left \langle \frac12 \Sigma^0 + \frac{\sqrt{3}}{2} \Lambda \Big| \overline{u} \gamma_\mu s \Big| \Xi^- \right\rangle = -\left \langle p \Big| \overline{u} \gamma_\mu s \Big| \frac12 \Sigma^0 - \frac{\sqrt{3}}{2} \Lambda \right\rangle,
\end{equation}
and
\begin{equation}
\left \langle \frac{\sqrt{3}}{2} \Sigma^0 - \frac12 \Lambda \Big| \overline{u} \gamma_\mu s \Big| \Xi^- \right \rangle = \left \langle p \Big| \overline{u} \gamma_\mu s \Big| \frac{\sqrt{3}}{2} \Sigma^0 + \frac12 \Lambda \right\rangle.
\end{equation}
\end{subequations}

Thus, in the limit of exact $SU(3)$ symmetry and neglecting isospin breaking, there are two relations among vector form factors, namely,
\begin{subequations}
\label{eq:sumr1}
\begin{equation}
\left[ f_1^{SU(3)} \right]_{\Xi^-\Sigma^0} + \sqrt{3} \left[ f_1^{SU(3)} \right]_{\Xi^-\Lambda} + \frac{1}{\sqrt{2}} \left[ f_1^{SU(3)} \right]_{\Sigma^-n} + \sqrt{3} \left[ f_1^{SU(3)} \right]_{\Lambda p} = 0,
\end{equation}
and
\begin{equation}
\sqrt{3} \left[ f_1^{SU(3)} \right]_{\Xi^-\Sigma^0} - \left[ f_1^{SU(3)} \right]_{\Xi^-\Lambda} - \sqrt \frac32 \left[ f_1^{SU(3)} \right]_{\Sigma^-n} + \left[ f_1^{SU(3)} \right]_{\Lambda p} = 0.
\end{equation}
\end{subequations}

Violations to relations (\ref{eq:sumr1}) are expected to occur due to flavor $SU(3)$ symmetry-breaking effects.

After some rearrangements, relations (\ref{eq:sumr1}) can be expressed, in the presence of SB, as
\begin{subequations}
\label{eq:srules}
\begin{equation}
\frac14 \left[ \frac{f_1}{f_1^{SU(3)}} \right]_{\Xi^-\Sigma^0} + \frac34 \left[ \frac{f_1}{f_1^{SU(3)}} \right]_{\Xi^-\Lambda} - \frac14 \left[ \frac{f_1}{f_1^{SU(3)}} \right]_{\Sigma^-n} - \frac34 \left[ \frac{f_1}{f_1^{SU(3)}} \right]_{\Lambda p} = \delta_1^{\mathrm{SB}},
\end{equation}
and
\begin{equation}
\frac34 \left[ \frac{f_1}{f_1^{SU(3)}} \right]_{\Xi^-\Sigma^0} - \frac34 \left[ \frac{f_1}{f_1^{SU(3)}} \right]_{\Xi^-\Lambda} + \frac34 \left[ \frac{f_1}{f_1^{SU(3)}} \right]_{\Sigma^-n} - \frac34 \left[ \frac{f_1}{f_1^{SU(3)}} \right]_{\Lambda p} = \delta_2^{\mathrm{SB}},
\end{equation}
\end{subequations}
where $\delta_k^{\mathrm{SB}}$ arise from SB and are formally of order $\mathcal{O}(\epsilon^2)$; hereafter, $\epsilon$ will be referred to as a measure of SB. The origin of these $\delta_k^{\mathrm{SB}}$ corrections will be explored in the context of the $1/N_c$ expansion of QCD.

\section{\label{sec:add}A brief review about the Ademollo-Gatto result}

By assuming that the vector currents and the electromagnetic current are members of the same unitary octet and that the breaking of the unitary symmetry behaves as the eighth component of an octet, M.~Ademollo and R.~Gatto set up an important theorem on the nonrenormalization for the strangeness-violating vector currents \cite{ag}.

Following Ademollo and Gatto, the $a$th component of the vector current $\mathcal{J}^a$ to first order in SB can be written as\footnote{The authors of Ref.~\cite{ag} inadvertently omitted to subtract singlet and octet pieces off in some terms. Conclusions remain unchanged, though.}
\begin{eqnarray}
\mathcal{J}^a + \epsilon \delta \mathcal{J}^a & = & a_0 \mathrm{Tr} (\overline{B}B\lambda^a) + b_0 \mathrm{Tr} (\overline{B} \lambda^a B) + \epsilon a \left[ \mathrm{Tr} (\overline{B}B \{\lambda^a,\lambda^8\}) - \frac18 \delta^{a8} \mathrm{Tr} (\overline{B}B \{\lambda^c,\lambda^c\}) \right] \nonumber \\
&  & \mbox{} + \epsilon b \left[ \mathrm{Tr} (\overline{B} \{\lambda^a,\lambda^8\} B) - \frac18 \delta^{a8} \mathrm{Tr} (\overline{B} \{\lambda^c,\lambda^c\} B) \right] \nonumber \\
&  & \mbox{} + \epsilon c [\mathrm{Tr} (\overline{B} \lambda^a B \lambda^8) - \mathrm{Tr} (\overline{B} \lambda^8 B \lambda^a)] + \epsilon g \mathrm{Tr} (\overline{B}B) \mathrm{Tr}(\lambda^a \lambda^8) \nonumber \\
&  & \mbox{} + \epsilon h \left[ \mathrm{Tr} (\overline{B} \lambda^a B \lambda^8) + \mathrm{Tr} (\overline{B} \lambda^8 B \lambda^a) - \frac14 \delta^{a8} \mathrm{Tr} (\overline{B} \lambda^c B \lambda^c) - \frac65 d^{a8c}d^{cgh} \mathrm{Tr} (\overline{B} \lambda^g B \lambda^h) \right], \label{eq:vcag}
\end{eqnarray}
where $B$ represents the baryon matrix, $\lambda^a$ denote the Gell-Mann matrices and $a_0$, $b_0$, \ldots, $h$ are coupling constants. The parameter $\epsilon$ is introduced to keep track of the number of times the perturbation enters; at the end of the calculation $\epsilon$ can be set to one without any loss of generality.

The electromagnetic current is defined in the usual way as
\begin{equation}
\mathcal{J}_{\mathrm{em}} = \mathcal{J}^Q \equiv \mathcal{J}^3 + \frac{1}{\sqrt{3}} \mathcal{J}^8,
\end{equation}
so the baryon charges, including first-order SB, are readily obtained from Eq.~(\ref{eq:vcag}). They read
\begin{equation}
Q_n + \epsilon \delta Q_n = -\frac23 a_0 - \frac23 b_0 + \frac{4}{3 \sqrt{3}} \epsilon a - \frac{8}{3 \sqrt{3}} \epsilon b + \frac{2}{\sqrt{3}} \epsilon c + \frac{2}{\sqrt{3}} \epsilon g + \frac{\sqrt{3}}{10} \epsilon h,
\end{equation}
\begin{equation}
Q_p + \epsilon \delta Q_p = -\frac23 a_0 + \frac43 b_0 + \frac{4}{3 \sqrt{3}} \epsilon a + \frac{4}{3 \sqrt{3}} \epsilon b - \frac{2}{\sqrt{3}} \epsilon c + \frac{2}{\sqrt{3}} \epsilon g - \frac{7\sqrt{3}}{10} \epsilon h,
\end{equation}
\begin{equation}
Q_{\Sigma^+} + \epsilon \delta Q_{\Sigma^+} = - \frac23 a_0 + \frac43 b_0 - \frac{8}{3 \sqrt{3}} \epsilon a + \frac{4}{3 \sqrt{3}} \epsilon b + \frac{2}{\sqrt{3}} \epsilon c + \frac{2}{\sqrt{3}} \epsilon g + \frac{\sqrt{3}}{10} \epsilon h,
\end{equation}
\begin{equation}
Q_{\Sigma^0} + \epsilon \delta Q_{\Sigma^0} = \frac13 a_0 + \frac13 b_0 - \frac{2}{3 \sqrt{3}} \epsilon a - \frac{2}{3 \sqrt{3}} \epsilon b + \frac{2}{\sqrt{3}} \epsilon g + \frac{\sqrt{3}}{10} \epsilon h,
\end{equation}
\begin{equation}
Q_{\Sigma^-} + \epsilon \delta Q_{\Sigma^-} = \frac43 a_0 - \frac23 b_0 + \frac{4}{3 \sqrt{3}} \epsilon a - \frac{8}{3 \sqrt{3}} \epsilon b - \frac{2}{\sqrt{3}} \epsilon c + \frac{2}{\sqrt{3}} \epsilon g + \frac{\sqrt{3}}{10} \epsilon h,
\end{equation}
\begin{equation}
Q_{\Xi^-} + \epsilon \delta Q_{\Xi^-} = \frac43 a_0 - \frac23 b_0 + \frac{4}{3 \sqrt{3}} \epsilon a + \frac{4}{3 \sqrt{3}} \epsilon b + \frac{2}{\sqrt{3}} \epsilon c + \frac{2}{\sqrt{3}} \epsilon g - \frac{7 \sqrt{3}}{10} \epsilon h,
\end{equation}
\begin{equation}
Q_{\Xi^0} + \epsilon \delta Q_{\Xi^0} = -\frac23 a_0 - \frac23 b_0 - \frac{8}{3 \sqrt{3}} \epsilon a + \frac{4}{3 \sqrt{3}} \epsilon b - \frac{2}{\sqrt{3}} \epsilon c + \frac{2}{\sqrt{3}} \epsilon g + \frac{\sqrt{3}}{10} \epsilon h,
\end{equation}
and
\begin{equation}
Q_\Lambda + \epsilon \delta Q_\Lambda = -\frac13 a_0 - \frac13 b_0 + \frac{2}{\sqrt{3}} \epsilon a + \frac{2}{\sqrt{3}} \epsilon b + \frac{2}{\sqrt{3}} \epsilon g + \frac{9\sqrt{3}}{10} \epsilon h,
\end{equation}
along with the isospin relation
\begin{equation}
\frac12 (Q_{\Sigma^+} + Q_{\Sigma^-}) = Q_{\Sigma^0}. \label{eq:isos}
\end{equation}

Solving the system of linear equations yields
\begin{equation}
a_0 = -\frac12, \quad b_0 = \frac12, \quad a = b = c = g = h = 0,
\end{equation}
which explicitly shows that first-order SB corrections to the electric current vanish. For the $|\Delta S|=1$ weak vector currents, the flavor index is $a = 4\pm i5$; according to the original assumption, first-order SB corrections also vanish for the vector current. This is in essence the celebrated result discovered by Ademollo and Gatto \cite{ag}.

\section{\label{sec:sur}A survey on the $1/N_c$ expansion of QCD}

In the large-$N_c$ limit, the baryon sector has a contracted $SU(2N_f)$ spin-flavor symmetry, where $N_f$ is the number of light quark flavors \cite{dm315,gs}. Corrections to the large-$N_c$ limit can be given in terms of
$1/N_c$-suppressed operators with well-defined spin-flavor transformation properties \cite{gs}; this yields the so-called $1/N_c$ expansion of QCD. For $N_f=3$, the lowest lying baryon states fall into a representation of the
spin-flavor group $SU(6)$. For $N_c=3$, the $\mathbf{56}$ dimensional representation is involved.

The $1/N_c$ expansion of any QCD operator transforming according to a given $SU(2)\times SU(N_f)$ representation can be written in terms of $n$-body operators $\mathcal{O}_n$ as \cite{djm95}
\begin{equation}
\mathcal{O}_\mathrm{QCD} = \sum_n c_{(n)} \frac{1}{N_c^{n-1}} \mathcal{O}_n,
\end{equation}
where the operator coefficients $c_{(n)}$ have power series expansions in $1/N_c$ beginning at order unity and the $\mathcal{O}_n$ are polynomials in the spin-flavor generators $J^k$, $T^c$, and $G^{kc}$, which can be written as 1-body quark operators acting on the $N_c$-quark baryon states, namely,
\begin{subequations}
\label{eq:su6gen}
\begin{eqnarray}
J^k & = & \sum_\alpha^{N_c} q_\alpha^\dagger \left(\frac{\sigma^k}{2}\otimes \openone \right) q_\alpha, \\
T^c & = & \sum_\alpha^{N_c} q_\alpha^\dagger \left(\openone \otimes \frac{\lambda^c}{2} \right) q_\alpha, \\
G^{kc} & = & \sum_\alpha^{N_c} q_\alpha^\dagger \left(\frac{\sigma^k}{2}\otimes \frac{\lambda^c}{2} \right) q_\alpha,
\end{eqnarray}
\end{subequations}
where $q_\alpha^\dagger$ and $q_\alpha$ are operators that create and annihilate states in the fundamental representation of $SU(6)$ and $\sigma^k$ and $\lambda^c$ are the Pauli spin and Gell-Mann flavor matrices, respectively. Because the baryon matrix elements of the spin-flavor generators (\ref{eq:su6gen}) can be taken as the values in the nonrelativistic quark model, this convention is usually referred to as the quark representation \cite{djm95}.

\section{\label{sec:bff}The baryon vector form factor in the $1/N_c$ expansion}

Now, let $V^{0c}$ denote the flavor octet baryon charge \cite{jen96},
\begin{equation}
V^{0c} = \left\langle B_2\left|\left(\overline{q} \gamma^0 \frac{\lambda^c}{2} q\right)_{\mathrm{QCD}}\right|B_1 \right\rangle,
\end{equation}
where the subscript QCD indicates that the quark fields are QCD quark fields rather than the quark creation and annihilation operators of the quark representation. $V^{0c}$ is spin-0 and a flavor octet, so it transforms as $(0,\mathbf{8})$ under $SU(2)\times SU(3)$; its matrix elements between $SU(6)$ symmetric states give the values of the leading vector form factor $f_1$. 

On general grounds, flavor SB in QCD is due to the strange quark mass $m_s$ and transforms as a flavor octet \cite{djm95}. To linear order in SB, the correction is obtained from the tensor product $(0,\mathbf{8})\times (0,\mathbf{8})$ so that the $SU͑(2͒)\times SU(3)$ representations involved are $(0,\mathbf{1})$, $(0,\mathbf{8})$, $(0,\mathbf{10+\overline{10}})$ and $(0,\mathbf{27})$ \cite{jl,rfm98}.

Let $V^c + \epsilon\delta V^c$ be the $(0,\mathbf{8})$ operator containing the most general first-order SB. Its $1/N_c$ expansion reads
\begin{eqnarray}
V^c + \epsilon \delta V^c & = & c_{(1)}^\mathbf{8} T^c + c_{(2)}^\mathbf{8} \frac{1}{N_c} \{J^r,G^{rc}\} + \epsilon N_c a_{(0)}^\mathbf{1} \delta^{c8} + \epsilon a_{(1)}^\mathbf{8} d^{c8e} T^e + \epsilon a_{(2)}^\mathbf{8} \frac{1}{N_c} d^{c8e} \{J^r,G^{re}\} \nonumber \\
&  & \mbox{} + \epsilon a_{(3)}^\mathbf{10+\overline{10}} \frac{1}{N_c^2} \left( \{T^c,\{J^r,G^{r8}\}\} - \{T^8,\{J^r,G^{rc}\}\} \right) \nonumber \\
&  & \mbox{} + \epsilon a_{(2)}^\mathbf{27} \frac{1}{N_c} \left[ \{T^c,T^8\} - \frac{N_f-2}{2N_f(N_f^2-1)} N_c(N_c+2N_f) \delta^{c8} - \frac{2}{N_f^2-1} \delta^{c8} J^2 - \frac{N_f-4}{N_f^2-4}(N_c + N_f) d^{c8e} T^e \right. \nonumber \\
&  & \mbox{\hglue1.9truecm} \left. - \frac{2N_f}{N_f^2-4} d^{c8e} \{J^r,G^{re}\} \right] \nonumber \\
&  & \mbox{} + \epsilon a_{(3)}^\mathbf{27} \frac{1}{N_c^2} \left[ \{T^c,\{J^r,G^{r8}\}\} + \{T^8,\{J^r,G^{rc}\}\} - \frac{4}{N_f(N_f+1)}(N_c+N_f) \delta^{c8} J^2 \right. \nonumber \\
&  & \mbox{\hglue1.9truecm} \left. - \frac{2}{N_f+2}(N_c + N_f) d^{c8e} \{J^r,G^{re}\} - \frac{2}{N_f+2} d^{c8e} \{J^2,T^e \} \right]. \label{eq:qfirst}
\end{eqnarray}

A few remarks are in order here. First, notice that the series has been truncated at the physical value $N_c=3$ so up to three-body operators should be retained. Second, the flavor singlet and octet components of the $\mathbf{27}$ operators have been explicitly subtracted off, so that only the truly flavor-$\mathbf{27}$ components remain. Third, the operator coefficients $c_{(n)}^\mathbf{8}$ come along with $n$-body operators given in the exact $SU(3)$ limit whereas the operator coefficients $a_{(n)}^\mathbf{rep}$ come along with $n$-body operators given in the representation $\mathbf{rep}$ which explicitly breaks flavor symmetry. And last but not least, higher-order operators are generated from the already existing ones by anticommuting with $J^2$. There is no need to include them because their contributions to the expansion can be accounted for by redefining the operator coefficients. Therefore, the series stands the way it is.

The matrix elements of the operator $V^Q+\epsilon \delta V^Q$ between $SU(6)$ symmetric states give the actual values of baryon charges $Q_B$ including first-order flavor SB. At the physical values $N_f=N_c=3$, the baryon charges read
\begin{equation}
Q_n + \epsilon \delta Q_n = -\frac13 c_{(2)}^\mathbf{8} + \sqrt{3} \epsilon a_{(0)}^\mathbf{1}-\frac{1}{\sqrt{3}} \epsilon a_{(1)}^\mathbf{8} - \frac{1}{2 \sqrt{3}} \epsilon a_{(2)}^\mathbf{8} + \frac{1}{3\sqrt{3}} \epsilon a_{(3)}^\mathbf{10+\overline{10}} - \frac{1}{20 \sqrt{3}} \epsilon a_{(2)}^\mathbf{27} - \frac{1}{30 \sqrt{3}} \epsilon a_{(3)}^\mathbf{27},
\end{equation}
\begin{equation}
Q_p + \epsilon \delta Q_p = c_{(1)}^\mathbf{8} + \frac12 c_{(2)}^\mathbf{8} + \sqrt{3} \epsilon a_{(0)}^\mathbf{1} + \frac{1}{3 \sqrt{3}} \epsilon a_{(2)}^\mathbf{8} - \frac{1}{3 \sqrt{3}} \epsilon a_{(3)}^\mathbf{10+\overline{10}} + \frac{7}{20 \sqrt{3}} \epsilon a_{(2)}^\mathbf{27} + \frac{7}{30 \sqrt{3}} \epsilon a_{(3)}^\mathbf{27},
\end{equation}
\begin{equation}
Q_{\Sigma^+} + \epsilon \delta Q_{\Sigma^+} = c_{(1)}^\mathbf{8}+\frac12 c_{(2)}^\mathbf{8} + \sqrt{3} \epsilon a_{(0)}^\mathbf{1} + \frac{1}{\sqrt{3}} \epsilon a_{(1)}^\mathbf{8} + \frac{1}{6 \sqrt{3}} \epsilon a_{(2)}^\mathbf{8} + \frac{1}{3\sqrt{3}} \epsilon a_{(3)}^\mathbf{10+\overline{10}} - \frac{1}{20 \sqrt{3}} \epsilon a_{(2)}^\mathbf{27} - \frac{1}{30 \sqrt{3}} \epsilon a_{(3)}^\mathbf{27},
\end{equation}
\begin{equation}
Q_{\Sigma^-} + \epsilon \delta Q_{\Sigma^-} = -c_{(1)}^\mathbf{8} - \frac16 c_{(2)}^\mathbf{8} + \sqrt{3} \epsilon a_{(0)}^\mathbf{1} - \frac{1}{\sqrt{3}} \epsilon a_{(1)}^\mathbf{8} - \frac{1}{2 \sqrt{3}} \epsilon a_{(2)}^\mathbf{8} - \frac{1}{3\sqrt{3}} \epsilon a_{(3)}^\mathbf{10+\overline{10}} - \frac{1}{20 \sqrt{3}} \epsilon a_{(2)}^\mathbf{27} - \frac{1}{30 \sqrt{3}} \epsilon a_{(3)}^\mathbf{27},
\end{equation}
\begin{equation}
Q_{\Sigma^0} + \epsilon \delta Q_{\Sigma^0} = \frac16 c_{(2)}^\mathbf{8} + \sqrt{3} \epsilon a_{(0)}^\mathbf{1} - \frac{1}{6 \sqrt{3}} \epsilon a_{(2)}^\mathbf{8} - \frac{1}{20 \sqrt{3}} \epsilon a_{(2)}^\mathbf{27} - \frac{1}{30\sqrt{3}} \epsilon a_{(3)}^\mathbf{27},
\end{equation}
\begin{equation}
Q_{\Xi^-} + \epsilon \delta Q_{\Xi^-} = -c_{(1)}^\mathbf{8} - \frac16 c_{(2)}^\mathbf{8} + \sqrt{3} \epsilon a_{(0)}^\mathbf{1} + \frac{1}{3 \sqrt{3}} \epsilon a_{(2)}^\mathbf{8} + \frac{1}{3 \sqrt{3}} \epsilon a_{(3)}^\mathbf{10+\overline{10}} + \frac{7}{20 \sqrt{3}} \epsilon a_{(2)}^\mathbf{27} + \frac{7}{30 \sqrt{3}} \epsilon a_{(3)}^\mathbf{27},
\end{equation}
\begin{equation}
Q_{\Xi^0} + \epsilon \delta Q_{\Xi^0} = -\frac13 c_{(2)}^\mathbf{8} + \sqrt{3} \epsilon a_{(0)}^\mathbf{1} + \frac{1}{\sqrt{3}} \epsilon a_{(1)}^\mathbf{8} + \frac{1}{6 \sqrt{3}} \epsilon a_{(2)}^\mathbf{8} - \frac{1}{3\sqrt{3}} \epsilon a_{(3)}^\mathbf{10+\overline{10}} - \frac{1}{20 \sqrt{3}} \epsilon a_{(2)}^\mathbf{27} - \frac{1}{30 \sqrt{3}} \epsilon a_{(3)}^\mathbf{27},
\end{equation}
\begin{equation}
Q_\Lambda + \epsilon \delta Q_\Lambda = -\frac16 c_{(2)}^\mathbf{8} + \sqrt{3} \epsilon a_{(0)}^\mathbf{1} + \frac{1}{6 \sqrt{3}} \epsilon a_{(2)}^\mathbf{8} - \frac{3}{20} \sqrt{3} \epsilon a_{(2)}^\mathbf{27} - \frac{1}{10} \sqrt{3} \epsilon a_{(3)}^\mathbf{27}.
\end{equation}
Although the isospin relation (\ref{eq:isos}) reduces by one the number of independent equations, it is straightforward to notice that the $a_{(n)}^\mathbf{27}$ operator coefficients are not independent. Indeed, a new coefficient $x_{(2)}^\mathbf{27}$ can be defined as
\begin{equation}
x_{(2)}^\mathbf{27} = a_{(2)}^\mathbf{27} + \frac23 a_{(3)}^\mathbf{27},
\end{equation}
so that the number of operator coefficients is also reduced by one.

Solving the system of linear equations yields
\begin{equation}
c_{(1)}^\mathbf{8} = 1, \qquad c_{(2)}^\mathbf{8} = 0, \qquad a_{(0)}^\mathbf{1} = 0, \qquad a_{(1)}^\mathbf{8} = 0, \qquad a_{(2)}^\mathbf{8} = 0, \qquad a_{(3)}^\mathbf{10+\overline{10}} = 0, \qquad x_{(2)}^\mathbf{27} = 0,
\end{equation}
which nicely reproduces the Ademollo-Gatto result. The coupling constants introduced in Eq.~(\ref{eq:vcag}) are related to the coefficients of the $1/N_c$ expansion (\ref{eq:qfirst}) at $N_c=3$ by
\begin{subequations}
\label{eq:coeffrel}
\begin{eqnarray}
&  & a_0 = -\frac12 c_{(1)}^\mathbf{8} + \frac{1}{12} c_{(2)}^\mathbf{8}, \\
&  & b_0 = \frac12 c_{(1)}^\mathbf{8} + \frac{5}{12} c_{(2)}^\mathbf{8}, \\
&  & a = -\frac14 a_{(1)}^\mathbf{8} + \frac{1}{24} a_{(2)}^\mathbf{8}, \\
&  & b = \frac14 a_{(1)}^\mathbf{8} + \frac{5}{24} a_{(2)}^\mathbf{8}, \\
&  & c = \frac16 a_{(3)}^\mathbf{10+\overline{10}}, \\
&  & g = \frac32 a_{(0)}^\mathbf{1}, \\
&  & h = - \frac16 x_{(2)}^\mathbf{27},
\end{eqnarray}
\end{subequations}
and they correspond to well-defined flavor representations.

Next, SB at second order can be incorporated into $V^c$. The $1/N_c$ expansion of this contribution reads
\begin{eqnarray}
\epsilon^2 \delta V^c & = & \epsilon^2 b_{(0)}^{\mathbf{1}}N_c d^{c88} \openone + \epsilon^2 b_{(1)}^{\mathbf{8}} \delta^{c8} T^8 + \epsilon^2 e_{(1)}^{\mathbf{8}} f^{c8e} f^{8eg} T^g + \epsilon^2 g_{(1)}^{\mathbf{8}} d^{c8e} d^{8eg} T^g \nonumber \\
&  & \mbox{} + \epsilon^2 h_{(1)}^{\mathbf{8}} (i f^{ceg} d^{8e8} T^g-i d^{ce8} f^{8eg} T^g-i f^{c8e} d^{eg8} T^g) + \epsilon^2 b_{(2)}^{\mathbf{8}} \frac{1}{N_c} \delta^{c8} \{J^r,G^{r8}\} + \epsilon^2 e_{(2)}^{\mathbf{8}} \frac{1}{N_c} f^{c8e} f^{8eg} \{J^r,G^{rg}\} \nonumber \\
&  & \mbox{} + \epsilon^2 g_{(2)}^{\mathbf{8}} \frac{1}{N_c} d^{c8e} d^{8eg} \{J^r,G^{rg}\} + \epsilon^2 h_{(2)}^{\mathbf{8}} \frac{1}{N_c} (i f^{ceg} d^{8e8} - i d^{ce8} f^{8eg} - i f^{c8e} d^{eg8}) \{J^r,G^{rg}\} \nonumber \\
&  & \mbox{} + \epsilon^2 b_{(2)}^{\mathbf{10+\overline{10}}}\frac{1}{N_c^2} d^{c8e} \left(\{T^e,\{J^r,G^{r8}\}\} - \{T^8,\{J^r ,G^{re}\}\} \right) \nonumber \\
&  & \mbox{} + \epsilon^2 b_{(2)}^{\mathbf{27}} \frac{1}{N_c} \left[ d^{c8e}\{T^e,T^8\} - \frac{N_f-4}{N_f^2-4}(N_c+N_f) d^{c8e} d^{8eg} T^g - \frac{2N_f}{N_f^2-4} d^{c8e} d^{8eg} \{J^r ,G^{rg}\} \right] \nonumber \\
&  & \mbox{} + \epsilon^2 b_{(3)}^{\mathbf{27}} \frac{1}{N_c^2} \left[ d^{c8e} \left( \{T^e,\{J^r,G^{r8}\}\} + \{T^8,\{J^r ,G^{re}\}\} \right) - \frac{2}{N_f+2}(N_c+N_f) d^{c8e} d^{8eg} \{J^r ,G^{rg}\} \right. \nonumber \\
&  & \mbox{\hglue2.0truecm} \left. - \frac{2}{N_f+2} d^{c8e} d^{8eg} \{J^2,T^g\} \right] \nonumber \\
&  & \mbox{} + \epsilon^2 b_{(3)}^{\mathbf{64}} \frac{1}{N_c^2} \left[ \{T^c,\{T^8,T^8\}\} -\frac{N_f-2}{N_f(N_f^2-1)}N_c(N_c+2N_f) \delta^{88} T^c - \frac{1}{2} f^{c8e}f^{8eg} T^g \right. \nonumber \\
&  & \mbox{\hglue2.0truecm} - \frac{N_f-4}{2(N_f^2-4)}(N_c+N_f) d^{c8e}\{T^e,T^8\} - \frac{N_f-4}{2(N_f^2-4)}(N_c+N_f) d^{88e} \{T^c,T^e\} - \frac{2}{N_f^2-1} \delta^{88} \{J^2,T^c\} \nonumber \\
&  & \mbox{\hglue2.0truecm} \left. - \frac{N_f}{N_f^2-4} d^{c8e}\{T^8,\{J^r,G^{re}\}\} - \frac{N_f}{N_f^2-4} d^{88e}\{T^c,\{J^r,G^{re}\}\} \right]. \label{eq:qsecond}
\end{eqnarray}

The matrix elements of the operator $\epsilon^2 \delta V^Q$ between $SU(6)$ symmetric states gives second-order SB effects to the baryon octet electric charges. For neutron, these corrections read
\begin{equation}
\epsilon^2 \delta Q_n = -\epsilon^2 b_{(0)}^\mathbf{1} + \frac12 \epsilon^2 b_{(1)}^\mathbf{8} + \frac{1}{12} \epsilon^2 b_{(2)}^\mathbf{8} + \frac13 \epsilon^2 g_{(1)}^\mathbf{8} + \frac16 \epsilon^2 g_{(2)}^\mathbf{8} - \frac19 \epsilon^2 b_{(3)}^\mathbf{10+\overline{10}} - \frac{1}{15} \epsilon^2 b_{(2)}^\mathbf{27} - \frac{2}{45} \epsilon^2 b_{(3)}^\mathbf{27} + \frac{13}{120} \epsilon^2 b_{(3)}^\mathbf{64},
\end{equation}
and similar expressions are found for the rest of the baryon charges. Again, the operator coefficients associated with the $\mathbf{27}$ representation are not independent, so a new coefficient $y_{(2)}^\mathbf{27}$ can be defined as
\begin{equation}
y_{(2)}^\mathbf{27} = b_{(2)}^\mathbf{27} + \frac23 b_{(3)}^\mathbf{27}.
\end{equation}

Only 8 out of 12 operator coefficients of expansion (\ref{eq:qsecond}) are involved in the seven expressions for the baryon charges. By using the important property that the electric charge remains unrenormalized to all orders in perturbation theory, the system can be solved in terms of one coefficient, namely,
\begin{subequations}
\label{eq:rest}
\begin{eqnarray}
&  & b_{(0)}^\mathbf{1} = b_{(2)}^\mathbf{8} = g_{(2)}^\mathbf{8} = y_{(2)}^\mathbf{27} = 0, \\
&  & g_{(1)}^\mathbf{8} = -\frac{5}{24} b_{(3)}^\mathbf{10+\overline{10}}, \\
&  & b_{(1)}^\mathbf{8} = -\frac{13}{72} b_{(3)}^\mathbf{10+\overline{10}}, \\
&  & b_{(3)}^\mathbf{64} = \frac52 b_{(3)}^\mathbf{10+\overline{10}}.
\end{eqnarray}
\end{subequations}

Thus, under the working assumptions, the baryon vector current is given to second-order in flavor SB in terms of, in principle, five nontrivial operator coefficients.

The matrix elements of $V^{4\pm i5}+\epsilon^2 \delta V^{4\pm i5}$ between $SU(6)$ baryon states yields the actual expressions for the leading vector form factors. For the observed processes, one has
\begin{equation}
\left[ \frac{f_1}{f_1^{SU(3)}} \right]_{\Lambda p} = 1 + \frac34 e_{(1)}^\mathbf{8} + \frac{1}{12} g_{(1)}^\mathbf{8} - \frac12 h_{(1)}^\mathbf{8} + \frac38 e_{(2)}^\mathbf{8} + \frac{1}{24} g_{(2)}^\mathbf{8} - \frac14 h_{(2)}^\mathbf{8} + \frac{1}{18} b_{(3)}^\mathbf{10+\overline{10}} - \frac{1}{10} y_{(2)}^\mathbf{27}, \label{eq:r1}
\end{equation}
\begin{equation}
\left[ \frac{f_1}{f_1^{SU(3)}} \right]_{\Sigma^-n} = 1 + \frac34 e_{(1)}^\mathbf{8} + \frac{1}{12} g_{(1)}^\mathbf{8} - \frac12 h_{(1)}^\mathbf{8} - \frac18 e_{(2)}^\mathbf{8} - \frac{1}{72} g_{(2)}^\mathbf{8} + \frac{1}{12} h_{(2)}^\mathbf{8} - \frac{1}{18} b_{(3)}^\mathbf{10+\overline{10}} - \frac{1}{30} y_{(2)}^\mathbf{27} + \frac{1}{30} b_{(3)}^\mathbf{64}, \label{eq:r2}
\end{equation}
\begin{equation}
\left[ \frac{f_1}{f_1^{SU(3)}} \right]_{\Xi^-\Lambda} = 1 + \frac34 e_{(1)}^\mathbf{8} + \frac{1}{12} g_{(1)}^\mathbf{8} - \frac12 h_{(1)}^\mathbf{8} + \frac18 e_{(2)}^\mathbf{8} + \frac{1}{72} g_{(2)}^\mathbf{8} - \frac{1}{12} h_{(2)}^\mathbf{8} + \frac{1}{18} b_{(3)}^\mathbf{10+\overline{10}} + \frac{1}{10} y_{(2)}^\mathbf{27}, \label{eq:r3}
\end{equation}
\begin{equation}
\left[ \frac{f_1}{f_1^{SU(3)}} \right]_{\Xi^-\Sigma^0} = 1 + \frac34 e_{(1)}^\mathbf{8} + \frac{1}{12} g_{(1)}^\mathbf{8} - \frac12 h_{(1)}^\mathbf{8} + \frac58 e_{(2)}^\mathbf{8} + \frac{5}{72} g_{(2)}^\mathbf{8} - \frac{5}{12} h_{(2)}^\mathbf{8} - \frac{1}{18} b_{(3)}^\mathbf{10+\overline{10}} + \frac{1}{30} y_{(2)}^\mathbf{27} + \frac{1}{30} b_{(3)}^\mathbf{64}. \label{eq:r4}
\end{equation}

Expressions (\ref{eq:r1})-(\ref{eq:r4}) are the most general ones which account for second-order SB effects in the hyperon semileptonic decay form factors. Substituting them into sum rules (\ref{eq:srules}) yields
\begin{equation}
\delta_1^{\mathrm{SB}} = \frac16 y_{(2)}^\mathbf{27}, \qquad \qquad \delta_2^{\mathrm{SB}} = -\frac16 b_{(3)}^\mathbf{10+\overline{10}} + \frac{1}{20} b_{(3)}^\mathbf{64}.
\end{equation}

Under restrictions (\ref{eq:rest}), the final form of sum rules (\ref{eq:srules}) become
\begin{subequations}
\label{eq:srulesfinal}
\begin{equation}
\frac14 \left[ \frac{f_1}{f_1^{SU(3)}} \right]_{\Xi^-\Sigma^0} + \frac34 \left[ \frac{f_1}{f_1^{SU(3)}} \right]_{\Xi^-\Lambda} - \frac14 \left[ \frac{f_1}{f_1^{SU(3)}} \right]_{\Sigma^-n} - \frac34 \left[ \frac{f_1}{f_1^{SU(3)}} \right]_{\Lambda p} = 0,
\end{equation}
and
\begin{equation}
\frac34 \left[ \frac{f_1}{f_1^{SU(3)}} \right]_{\Xi^-\Sigma^0} - \frac34 \left[ \frac{f_1}{f_1^{SU(3)}} \right]_{\Xi^-\Lambda} + \frac34 \left[ \frac{f_1}{f_1^{SU(3)}} \right]_{\Sigma^-n} - \frac34 \left[ \frac{f_1}{f_1^{SU(3)}} \right]_{\Lambda p} = -\frac{1}{24} b_{(3)}^\mathbf{10+\overline{10}}.
\end{equation}
\end{subequations}

These findings are remarkable. Sum rule (\ref{eq:srulesfinal}a) is valid in the presence of second-order SB whereas (\ref{eq:srulesfinal}b) gets corrections mainly from the $\mathbf{10}+\overline{\mathbf{10}}$ flavor representation.

Different methods have been used to evaluate SB effects to vector form factors. One of them is baryon chiral perturbation theory (BChPT) in the works by Krause \cite{krause} and Anderson and Luty \cite{and}. The former presented the calculation in relativistic BChPT to order $\mathcal{O}(p^2)$ in the chiral expansion whereas the latter used heavy baryon chiral perturbation theory (HBChPT) up to (partially complete) order $\mathcal{O}(p^3)$. Both calculations included as baryonic degrees of freedom only the spin-1/2 octet. Later on, Villadoro \cite{villa} used HBChPT with both octet and decuplet baryon degrees of freedom and included (partially) up order $\mathcal{O}(p^3)$ corrections corresponding to subleading in $1/M_B$ terms. In the context of covariant BChPT, two schemes are representative. The first one used infrared regularization in the work by Lacour, Kubis, and Meissner \cite{meiss} and the second one used the extended-on-mass-shell (EOMS) renormalization scheme in the work by Geng, Camalich, and Vicente-Vacas \cite{geng}. Both works performed calculations to order $\mathcal{O}(p^3)$; while the former included only octet baryons as active degrees of freedom, the latter did include both octet and decuplet baryons. A more recent calculation within large-$N_c$ baryon chiral perturbation theory to order $\mathcal{O}(p^2)$ has been presented in Ref.~\cite{fmg}. In this approach loop graphs with octet and decuplet intermediate states are systematically incorporated into the analysis because both spin-1/2 and spin-3/2 baryons together form an irreducible representation of spin-flavor symmetry.

Sum rule (\ref{eq:srulesfinal}a) is analytically fulfilled by the $f_1/f_1^{SU(3)}$ ratios given in Eqs.~(65)--(68) of Ref.~\cite{fmg}; it is also satisfied when the decuplet fields are not explicitly retained
in the effective theory but integrated out. In consequence, this sum rule is fulfilled by all the expressions for the $f_1/f_1^{SU(3)}$ ratios obtained within (heavy) baryon chiral perturbation theory to order $\mathcal{O}(p^2)$ of Refs.~\cite{krause,villa,meiss,geng}. The only exception is found in Ref.~\cite{and} where there is a wrong sign in one of the loop diagrams. Terms of subleading order in $1/M_B$ computed in Refs.~\cite{villa,meiss,geng} also satisfy sum rule (\ref{eq:srulesfinal}a).

As for sum rule (\ref{eq:srulesfinal}b), its right-hand side yields
\begin{equation}
\frac32 D^2 \left[H(m_K,m_\eta)-H(m_\pi,m_K)\right] - \frac38 \mathcal{C}^2 \left[K(m_K,m_\eta,\Delta)-K(m_\pi,m_K,\Delta)\right], \label{eq:sr2}
\end{equation}
where $D$ and $\mathcal{C}$ are the usual $SU(3)$ invariant couplings, the functions $H(m_1,m_2)$ and $K(m_1,m_2,\Delta)$ come from loop integrals and $\Delta$ is the decuplet-octet baryon mass difference \cite{fmg}. Numerically, the quantities between square parentheses in Eq.~(\ref{eq:sr2}) are $0.056$ and $0.029$, respectively. As a side remark, the large-$N_c$ cancellations between loop diagrams in Eq.~(\ref{eq:sr2}) are guaranteed to occur as a consequence of the contracted spin-flavor symmetry which is present in the $N_c\to \infty$ limit.

\section{\label{sec:sum}Closing remarks}

To close this paper, it should be stressed that two sum rules for leading vector form factors in hyperon semileptonic decays, (\ref{eq:srulesfinal}), have been obtained by exploiting only the general properties shared by the weak currents and the electromagnetic current under flavor $SU(3)$ symmetry. The $1/N_c$ expansion of QCD has been used to evaluate the group theoretic structure of the corrections. The sum rules have been tested against symmetry-breaking patterns obtained within (heavy) baryon chiral perturbation theory to order $\mathcal{O}(p^2)$ are they are well fulfilled. A full test beyond that chiral order currently is not possible because the existing calculations present partially complete or rather contradictory results among them. In the near future, lattice QCD could also be used to explore these sum rules; both the $1/N_c$ and SB hierarchies should be manifest.

\section{acknowledgments}

The authors thank J.L.\ Goity for helpful communications. This work has been partially supported by Consejo Nacional de Ciencia y Tecnolog{\'\i}a (M\'exico) and Fondo de Apoyo a la Investigaci\'on (Universidad Aut\'onoma de San Luis Potos{\'\i}).

\end{document}